\documentclass[a4paper,twocolumn,english,aps,prl]{revtex4}
\usepackage[T1]{fontenc}
\usepackage[latin1]{inputenc}
\usepackage{float}
\usepackage{graphicx,color}

\makeatletter

\providecommand{\LyX}{L\kern-.1667em\lower.25em\hbox{Y}\kern-.125emX\@}

\usepackage{babel}
\makeatother
\begin{document}

\preprint{This line only printed with preprint option}

\title{Dispersion interaction between two atoms in electromagnetic fields}

\author{Y. Sherkunov}

\affiliation{Department of Physics, University of Warwick, Coventry, CV4 7AL,
UK}

\begin{abstract}
We present a new theory of atom-atom dispersion interaction in the presence of electromagnetic fields. The theory takes into account the absorption and emission of virtual photons leading to the resonance contributions to the interaction potential in the case of non-equilibrium dynamics.

\end{abstract}
\maketitle


Recent progress in experimental methods \cite{Mohideen98,Decca03, Bordag01, Lamoreaux05} enabling one to measure the thermal component of the Casimir-Lifshitz force \cite{Obrecht07}, stimulated a new wave of theoretical works in Casimir, Casimir-Polder (CP), and van der Waals dispersion forces at the presence of electromagnetic field \cite{Cohen02,Henkel02, Antezza05, Antezza08}. The problem of the Casimir-Lifshitz interaction between real metals is still not resolved. Although, the interaction is described by the famous Lifshitz theory, the theoretical predictions dramatically depend of the model describing the real metal (see for example \cite{Brevik05, Bezerra06, Mostepanenko06} and references therein). Another problem to be clarified is the non-equilibrium CP interaction between a single atom and a metal (or dielectric) plate if the temperature of the plate differs from the temperature of the atomic gas. This force was recently measured with the help of spectroscopy of an atom interacting with the dielectric slub \cite{Chaves07}, but the agreement with the theoretical predictions has not been achieved \cite{Chaves07}.  

The theoretical treatment of the CP interaction under non-equilibrium conditions is based either on the linear-responce (or Lifshitz) theory \cite{Antezza05, Henkel02, McLachlan63}, or macroscopic quantum electrodynamics \cite{Gorza06}. As it has been demonstrated by Buhmann and Scheel \cite{Buhmann08} one should distinguish two different problems. If the atom is thermalized and coupled to its thermal bath, the both approaches lead to the same result. If the atom is not thermalized, the results of the linear-responce theory could underestimate the the CP force \cite{Buhmann08}. The macroscopic QED approach takes into account the possible absorption and spontaneous/stimulated emission of a thermal photon by the interacting atom. These contributions result in the resonance enhancement of the CP force even if the atom is in its ground state.   For case of thermalized atom, this terms cancel out and one arrives at the predictions of Lifshitz theory \cite{Lifshitz}.

To understand the physical mechanism of the CP interaction under non-equilibrium conditions it is important to have a clear picture of the simplest case - the dispersion interaction between two atoms in electromagnetic fields. Milonni and Smith treated such kind of interaction \cite{MilonniSmith96} with the help of source theory \cite{MilonniShih92}. They showed that the interaction potential between two atoms in electromagnetic field could be obtained from the vacuum field potential by the simple substitution $\hbar \omega_{\mathbf{k}}/2 \rightarrow \hbar \omega_{\mathbf{k}}\left(N_{\mathbf{k}\lambda}+1/2\right)$ \cite{MilonniSmith96}, where $\omega_\mathbf{k}$ is the energy of the photon of $\{\mathbf{k},\lambda\}$ state, with $\mathbf{k}$ and $\lambda$ the wave vector and the polarization index of EM field, $N_{\mathbf{k}\lambda}$ is the number of photons in the state $\{\mathbf{k}\lambda\}$. Deriving their formula, Milonni and Smith took into account the scattering of electromagnetic field on atoms but they ignored the virtual absorption and/or spontaneous/stimulated emission of photons by the atoms. Obviously, this approach is valid for the case of equilibrium due to the detailed balance between the radiation and the state of atoms. But if the atoms and the radiation are not in the equilibrium their result is in contradiction with the recent theoretical results obtained for CP force between an atom and a dielectric slub \cite{Buhmann08, Gorza06} taking into account virtual absorption and emission of photons. 

Recent theoretical and experimental progress in non-equilibrium Casimir physics inspired us to reexamine the interaction between two atoms in electromagnetic field. To calculate the interaction potential we use a QED method based on the Keldysh Green function technique \cite{Sherkunov07}. We start with the derivation of the general formula for the interaction between a ground state atom and a dielectric medium in electromagnetic field.  To simplify the calculations we suppose that the number of photons of the external EM field is negligible at the transition energies of the ground-state atom. This simplification enables us to treat the atom as a ground-state one during the interaction and calculate the interaction potential as the energy shift of the atomic level. Then we consider the interaction of a ground-state atom with a medium at non-zero temperatures, provided the atom is not thermalized. We rederived the Buhmann-Scheel formula \cite{Buhmann08}. Next, we consider the interaction between two atoms in electromagnetic field and demonstrate that the absorption or spontaneous emission of photons play major role for non-equilibrium case resulting in resonance terms of the interaction potential. We show that the dispersion force for an excited atom and a ground-state one is enhanced by factor $N(\omega_B)+1$, where $\omega_B$ is the transition frequency for the excited atom, compared to the Power and Thirunamachandran result for electromagnetic vacuum \cite{Power93, Power95}. If both atoms are in their ground states, the difference between the vacuum case and the case of external field is more dramatic. The presence of the electromagnetic field results in the resonance force, which drops like $R^{-2}$ with the distance between the atoms $R$ at the retarded regime ($R>>2\pi/k$).  We discuss the validity of the Milonni-Smith formula.

  
We start with the CP interaction between a single ground-state atom (atom $A$) at a position $\mathbf{R}_{A}$ and an arbitrary dielectric medium at the presence of electromagnetic field. If the number of photons at the transition frequencies of the atom is negligible, the atom does not change its internal state. In this case we can treat the CP potential as the energy shift of the ground level of the atom $U=\Delta\epsilon_{A}$ \cite{CasimirPolder}. The Hamiltonian of the system atom $A$ - electromagnetic field - medium reads
\begin{eqnarray}
 H&=&\sum_i\epsilon_i b_i^{\dagger}b_i+\sum_{\mathbf{k}\lambda}k\left(\alpha_{\mathbf{k}\lambda}^{\dagger}\alpha_{\mathbf{k}\lambda}+1/2\right)+H_{med}\nonumber\\
& & -\int \psi^{\dagger}(X)\mathbf{d}\mathbf{E}(X)\psi(X)d\mathbf{r}
\end{eqnarray}
Where $\epsilon_i$ is the bare energy of atom $A$ at state $i$, $b_i$ is the annihilation operator of this state, $\mathbf{k}$ is the wave vector of electromagnetic field, $\lambda$ is the polarization index, $\alpha_{\mathbf{k}\lambda}$ is the annihilation operator of state $\{\mathbf{k},\lambda\}$ of the electromagnetic field, $H_{med}$ is the Hamiltonian of the medium interacting with the electromagnetic field, $\psi(X)=\sum_i\phi_i(\mathbf{r},\mathbf{R}_A)b_i$ with $\phi$ wavefunction of the atom $A$, $X=\{\mathbf{r},t\}$, $\mathbf{E}$ is the operator of electromagnetic field. $\mathbf{d}$ is the dipole moment of the electron.  

Following the earlier treatment for a dipole interaction of a ground state atom with a dielectric medium in electromagnetic vacuum \cite{Sherkunov07}, we generalize our previous result to the case of electromagnetic field surrounding the atom and the medium \cite{Sherkunov05}. 
The retarded Green function of atom $A$  obeys the Dyson equation. In derivation we followed the standard Keltysh method for the electromagnetic field and a modified one for the atoms \cite{Sherkunov05,Sherkunov07}. Treating the atom, we directly implemented Wick's theorem for a single atom making use of the fact that the normal products of the atomic operators of all orders but the second one are zero. The normal product of the second order is just the density matrix of the atom $A$ \cite{Sherkunov05}.    In energy representation the Dyson equation reads
\begin{eqnarray}
G_r(\omega,\mathbf{r},\mathbf{r}')=G_r^0(\omega,\mathbf{r},\mathbf{r}')\nonumber\\
+\int G_r^0(\omega,\mathbf{r},\mathbf{r}_1)M_r(\omega,\mathbf{r}_1,\mathbf{r}_2)G_r(\omega,\mathbf{r}_2,\mathbf{r}')d\mathbf{r}_1d\mathbf{r}_2
\label{Dyson}
\end{eqnarray}
where $G_r^0$ is the retarded Green function for a non-interacting atom, $M_r$ is the mass operator \cite{Sherkunov05}
\begin{equation}
 M_r(X,X')=id^\nu d^{\nu'}G_r(X,X')D^{\nu \nu'}(X',X)
\end{equation}
Where $D$ is the casual photon Green function
\begin{equation}
 D^{\nu\nu'}(X,X')=-i\langle \hat{T}E^{\nu'}(X')E^{\nu}(X)\rangle
\end{equation}
$\hat{T}$ is the time-ordering operator
Neglecting the transition of the atom $A$ to its exciting state and using the pole approximation we arrive at the solution of the Dyson equation (\ref{Dyson})
\begin{equation}
 G_r(\omega)=\sum_{i}\frac{\phi_i(\mathbf{r},\mathbf{R}_A)\phi_i^*(\mathbf{r}',\mathbf{R}_A)}{\omega-\epsilon_i-M_r^{ii}(\epsilon_i)}
\end{equation}
with $M_r^{ii}=\int\phi_i^*(\mathbf{r})M_r(\mathbf{r},\mathbf{r}')\phi_i(\mathbf{r}')d\mathbf{r}d\mathbf{r}'$. 
Thus, the energy shift of the ground level of atom $A$ is $U_A=ReM_r^{ii}$. 
Following the calculations of \cite{Sherkunov05,Sherkunov07}, we obtain
\begin{equation}
 U_{A}=-Re\frac{i}{2\pi}\int_0^\infty\alpha_{A}^{\nu'\nu}(\omega)D^{\nu\nu'}(\omega,\mathbf{R}_{A},\mathbf{R}_{A})d\omega
\label{main}
\end{equation}
This formula generalizes the one \cite{Sherkunov07} obtained for the interaction of an atom with a dielectric medium at zero temperature if the electromagnetic field is in its vacuum state. Now we include the external EM field.
The polarizability of the atom at the internal state $k$ is given by the standard formula (we put $\hbar=1, c=1$ throughout the paper)
\begin{equation}
 \alpha_{Ak}^{\nu'\nu}(\omega)=\sum_j\left( \frac{d_{kj}^{\nu'}d_{jk}^\nu}{\omega_{jk}-\omega-i0}+\frac{d_{jk}^{\nu'}d_{kj}^\nu}{\omega_{jk}+\omega+i0}\right) 
\label{polarizability}
\end{equation}
with $\omega_{kj}$ transition frequency of the atom from state $k$ to state $j$, $d_{jk}^{\nu}$ the $\nu$-th projection of electric dipole matrix element between states 
$j$ and $k$, $i0$ describes the analytical properties of the polarizability.

To calculate the Green tensor we use the standard Keldysh technique \cite{Keldysh, LandauX}. First we notice that the casual Green tensor can be written as a sum of the retarded Green tensor and the Green tensor $D_{12}$ \cite{Keldysh, LandauX}
\begin{equation}
 D=D_r+D_{12}
\label{sum}
\end{equation}
where 
\begin{equation}
 D_{12}^{\nu\nu'}(X,X')=-i\langle \hat{E}^{\nu'}(X')\hat{E}^{\nu}(X)\rangle
\label{desmatr}
\end{equation}
Substituting (\ref{desmatr}) and (\ref{sum}) into (\ref{main}) we find
\begin{eqnarray}
& & U_{A} = -Re\frac{i}{2\pi}\int_0^\infty\alpha_{A}^{\nu'\nu}(\omega)D_{r}^{\nu\nu'}(\omega,\mathbf{R}_{A},\mathbf{R}_{A})d\omega \nonumber\\
& &-Re\frac{i}{2\pi}\int_0^\infty\alpha_{A}^{\nu'\nu}(\omega)D_{12}^{\nu\nu'}(\omega,\mathbf{R}_{A},\mathbf{R}_{A})d\omega
\label{sep}
\end{eqnarray}
The first term of the Eq.(\ref{sep}) is the standard interaction potential for the ground-state atom interacting with the vacuum electromagnetic field at the presence of an arbitrary dielectric body \cite{WylieSipe}. The second term describes the interaction of the atom with the photons of the external electromagnetic field or the photons radiated by the media. 

For zero temperature the density matrix $D_{12}$ describes only spontaneous emission by the medium and Ex. (\ref{sep}) results in Ex. (41) of \cite{Sherkunov07}.

As the first example we consider a ground-state atom embedded in the system electromagnetic field - dielectric medium at thermal equilibrium at temperature $T$. We suppose that the atom does not change its initial state, it means that the time scales are short compared to the inverse ground-state heating rates of the atom $\Gamma_{kj}^{-1}$. This situation
was recently considered by Bushmann and Scheel \cite{Buhmann08}. 
The casual Green tensor of the thermal electromagnetic field is \cite{LandauIX}
\begin{eqnarray}
& & D(\omega,\mathbf{R}_{A},\mathbf{R}_{A}) = ReD_{r}(\omega,\mathbf{R}_{A},\mathbf{R}_{A}) \nonumber\\
& & +i\coth(\omega/2T)ImD_{r}(\omega,\mathbf{R}_{A},\mathbf{R}_{A})
\end{eqnarray}
It can be rewritten as
\begin{equation}
D=(2N(\omega)+1)D_{r}-2N(\omega)ReD_{r}
\end{equation}
where $N(\omega)$ is the occupation number of the photons at frequency $\omega$
Using Ex. (\ref{main}), we find for an isotopic atom
\begin{eqnarray}
& & U_{A} = T\sum_{m=0}^{\infty}(1-\frac{1}{2}\delta_{m0})\alpha_{A}^{\nu'\nu}(i\xi_{m})D_r^{\nu\nu'}(i\xi_{m})\nonumber\\
& & -1/3\sum_{j}N(\omega_{jk})|d_{kj}|^2ReD_r(\omega_{jk})\theta(\omega_{jk})
\end{eqnarray}
where $\xi_m=2\pi mT$ is the Mazubara frequency.
This result coincides with the Ex. (25) of \cite{Buhmann08}. The first term describe the standard Lifshitz result for the interaction of a \textit{thermalized} atom with the medium at thermal equilibrium. The second term, as it was pointed out by Buhmann and Scheel, is the resonance contribution to the Casimir-Polder force due to absorption of thermal photons by the atom, which is not thermalized \cite{Buhmann08}.

Let a ground-state atom $A$ be at a position $\mathbf{R}_{A}$ and atom $B$ at a position $\mathbf{R}_{B}$. Atom $B$ can be either in excited or ground state. For simplicity we suppose  that the atoms are exposed to isotropic unpolarized electromagnetic field, \textit{i.e.} $N_{\mathbf{k}\lambda}$ depends only on $|\mathbf{k}|=\omega$ , $N_{\mathbf{k}\lambda}=N(\omega)$. We suppose that $N(\omega_{jk}^A)=0$ to simplify the calculation. 
We will attack the problem perturbatively. The density matrix $D_{12}$ can be calculated with the help of Keldysh technique \cite{Keldysh, LandauX}. According to (\ref{main}), we should integrate the Green tensor with respect to positive frequencies $\omega >0$
\begin{eqnarray}
D_{12}(X,X')=D_{12}^0\nonumber\\
+\int D_r^0(X,X_1)\Pi_r(X_1,X_2)D_{12}^0(X_2,X')dX_1dX_2\nonumber\\
+\int D_{12}^0(X,X_1)\Pi_a(X_1,X_2)D_{a}^0(X_2,X')dX_1dX_2\nonumber\\
-\int D_{r}^0(X,X_1)\Pi_{12}(X_1,X_2)D_{a}^0(X_2,X')dX_1 dX_2
\label{d12}
\end{eqnarray}
Here we suppressed the tensor indices $\nu$. $D^0$  means free photon Green tensor. $X=\{\mathbf{r},t\}$
The polarization operators $\Pi$ describe interaction of electromagnetic field with atom $B$ which is in its initial state $n$ \cite{Sherkunov07}
\begin{eqnarray}
 \Pi_r^{\nu\nu'}(\omega,\mathbf{r},\mathbf{r}')=-\alpha_B^{\nu\nu'}(\omega)\delta(\mathbf{r}-\mathbf{r}')=\Pi_a^{\nu\nu'*}(\omega,\mathbf{r}',\mathbf{r})\nonumber\\
\Pi_{12}^{\nu\nu'}(\omega,\mathbf{r},\mathbf{r}')=\sum_m2\pi id_{nm}^{\nu B}d_{mn}^{\nu'B}\delta(\omega-\omega_{nm})\delta(\mathbf{r}-\mathbf{r}')
\label{pi}
\end{eqnarray}
The polarization operators $\Pi_{r}$ and $\Pi_{a}$ describe elastic scattering of light on the atom B, while $\Pi_{12}$ describes the downward transition of the atom as a result of spontaneous or stimulated emission. 
For positive frequencies ($\omega>0$),the free field Green tensor $D_{12}^0$ we can represented as
\begin{equation}
 D_{12}^0(\omega,\mathbf{r},\mathbf{r}')=N(\omega)(D_r^0(\omega,\mathbf{r},\mathbf{r}')-D_a^0(\omega,\mathbf{r},\mathbf{r}'))
\label{1}
\end{equation}
The formula (\ref{d12}) along with (\ref{pi}) and (\ref{1}) yield
\begin{eqnarray}
D_{12}(\omega ,\mathbf{R}_A , \mathbf{R}_A)=D_{12}^{0}(\omega , \mathbf{R}_A , \mathbf{R}_A)\nonumber\\
+2N(\omega)D_{r}^{0}(\omega , \mathbf{R}_A , \mathbf{R}_B)\Pi_r(\omega)D_{r}^{0}(\omega , \mathbf{R}_B , \mathbf{R}_A)\nonumber\\
-2N(\omega)ReD_{r}^{0}(\omega , \mathbf{R}_A , \mathbf{R}_B)\Pi_r(\omega)D_{r}^{0}(\omega , \mathbf{R}_B , \mathbf{R}_A)\nonumber\\
+N(\omega)D_{r}^{0}(\omega , \mathbf{R}_A , \mathbf{R}_B)\Pi_{21}(\omega)D_{a}^{0}(\omega , \mathbf{R}_B , \mathbf{R}_A)\nonumber\\
-(N(\omega)+1)D_{r}^{0}(\omega , \mathbf{R}_A , \mathbf{R}_B)\Pi_{12}(\omega)D_{r}^{0}(\omega , \mathbf{R}_B , \mathbf{R}_A)
\label{2}
\end{eqnarray}
The polarization operator $\Pi_{21}$ describes the absorption of a photon by atom $B$ and an upward transition of the atom $B$.
\begin{eqnarray}
 \Pi_{21}^{\nu\nu'}(\omega,\mathbf{r},\mathbf{r}')=\sum_m2\pi id_{nm}^{\nu B}d_{mn}^{\nu'B}\delta(\omega-\omega_{mn})\delta(\mathbf{r}-\mathbf{r}')
\end{eqnarray}

The first term of the r.h.s of Eq.(\ref{2}) describes the free electromagnetic field. This term results in the optical Stark shift of the energy levels of atom $A$ and we omit this term. The second and the third terms describe the elastic scattering of light on atom $B$. The fourth term is responsible for absorption of a photon by atom $B$ and the last term is for the spontaneous and stimulated radiation of a photon by atom $B$.  
For isotopic atom $B$ prepared in a mixed state with the probability to find the atom in $n$-th state $p_n$ we obtain
\begin{eqnarray}
U_A=Re\frac{i}{\pi}\int_{0}^{\infty}d\omega\left(N(\omega)+1/2\right)\alpha_A (\omega) \alpha_B(\omega)\nonumber\\ 
\times(D_r^0(\omega,\mathbf{R}_A\mathbf{R}_B))^2\nonumber\\
+\frac{1}{3}Re\sum_{mn}|d_{mn}^B|^2\alpha_A(\omega_{mn})|D_r^0(\omega_{mn},\mathbf{R}_A,\mathbf{R}_B)|^2\nonumber\\
\times\left[p_n N(\omega_{mn})\theta(\omega_{mn})-p_n\left(N(\omega_{nm})+1\right)\theta(\omega_{nm})\right]
\label{3}
\end{eqnarray}
The first term of the Eq. (\ref{3}) coincides with the one obtained by Milonni and and Smith for interaction between two atoms in electromagnetic field using the source theory \cite{MilonniSmith96}. The second term describes the absorption of a photon by atom $B$ (first term in the square brackets) as well as spontaneous and stimulated emission by atom $B$ (second term in the square brackets). The second term is resonant. It could be easily checked by substitution of the polarizability of atom $A$ (Ex.(\ref{polarizability})). Let us consider a case of thermal equilibrium. The probability to find an atom in its $n$-th state $p_n$ is given by the Boltzmann distribution, while the number of photons obey the Bose distribution. One can easily check that in equilibrium, the second term is zero due to detailed balance, and we find that the equation (\ref{3}) coincides with the Lifshitz formula \cite{Lifshitz}.

 Thus we arrive at a conclusion that the Milonni formula describes only the interaction between two atoms at equilibrium. But for the non-equilibrium case, if one or both atoms are not thermalized, the Milonni formula, which does not take into account the absorption and emission of photons, underestimates the interaction potential between two atoms in the electromagnetic field even if the atoms are in their ground states. 

Now to obtain quantitative results, we substitute the explicite expression for the retarded Green tensor \cite{LandauIX}
\begin{eqnarray}
& &  D_r^{\nu\nu'0}(\omega,\mathbf{r},\mathbf{r}')= \nonumber\\
& &\times\omega^2\left[\delta_{\nu\nu'}\left(1+\frac{i}{\omega R}-\frac{1}{(\omega R)^2}\right)\right.\nonumber\\
& &\left. +s^{\nu}s^{\nu'}\left(\frac{3}{(\omega R)^2}-\frac{3i}{\omega R}-1\right)\right]\frac{\exp(i\omega R)}{R},
\end{eqnarray}
where $R=|\mathbf{r}-\mathbf{r}'|$ and $s^{\nu}=(r-r')^{\nu}/|\mathbf{r}-\mathbf{r}'|$, into (\ref{3})

For the non-resonance and the resonance parts of the potential we find
\begin{eqnarray}
& &U_{Anr}=-\frac{1}{\pi R^2}\int_0^\infty\alpha_A(iu)\alpha_B(iu)u^4\left(1+\frac{2}{uR}\right.\nonumber\\
& &+\left.\frac{5}{(uR)^2}+\frac{6}{(uR)^3}+\frac{3}{(uR)^4}\right)\nonumber\\
& &\times\left(2N(iu)+1\right)\exp[-2iuR]du\\
\label{5}
& & U_{Ares}=\frac{4}{9R^2}\sum_{mnj}\frac{|d_{kj}^A|^2|d_{mn}^B|^2\omega_{kj(A)}\omega_{mn(B)}^4}{\omega_{kj(A)}^2-\omega_{mn(B)}^2}\nonumber\\
& &\times\left(1+\frac{1}{(\omega_{kj(A)}R)^2}+\frac{3}{(\omega_{kj(A)}R)^4}\right)\nonumber\\
& &\times\left[p_nN(\omega_{mn(B)})\theta(\omega_{mn(B)})\right.\nonumber\\
& &\times\left.-p_n(N(\omega_{nm(B)})+1)\theta(\omega_{nm(B)})\right].
\label{4}
\end{eqnarray}

If the electromagnetic field is in its vacuum state ($N(\omega)=0)$) the first term in the square brackets of (\ref{4}) is zero. The second term is not equal to zero only if atom $B$ is excited. In this case we obtain the result by Power and Thirunamachandran \cite{Power93, Power95}.
\begin{eqnarray}
& &U_{Ares}=-\frac{4}{9R^2}\sum_{mj}\frac{|d_{kj}^A|^2|d_{mn}^B|^2\omega_{kj(A)}\omega_{mn(B)}^4}{\omega_{kj(A)}^2-\omega_{mn(B)}^2}\nonumber\\
& &\times\left(1+\frac{1}{(\omega_{kj(A)}R)^2}+\frac{3}{(\omega_{kj(A)}R)^4}\right)\theta(\omega_{nm(B)}).
\end{eqnarray}
For the limit $R>>2\pi/k$, when  the retardation effects are significant, the potential drops with the distance between the atoms as $U_{Ares}\propto R^{-2}$ (compare to the non-resonance contribution which is given by Casimir-Polder formula $U_{Anr}\propto R^{-7}$), this weird behavior of the potential was discussed by Power and Thirunamachandran \cite{Power93, Power95}. (see also \cite{Sherkunov07}, where the interaction potential between an excited atom and a ground-state one embedded in a dielectric medium was calculated for vacuum state of electromagnetic field). In the presence of electromagnetic field, the resonance contribution to the interaction between a ground-state atom and an excited one is enhanced by factor $N(\omega_B)+1$ compared to the interaction potential in electromagnetic vacuum.
 
If both atoms are in their ground state, we should take into account only the first term in square brackets of (\ref{4})
For the retarded regime ($R>>2\pi/k$)
\begin{eqnarray}
& & U_{Ares}=\frac{4}{9R^2}\sum_{mj}\frac{|d_{kj}^A|^2|d_{mn}^B|^2\omega_{kj(A)}\omega_{mn(B)}^4}{\omega_{kj(A)}^2-\omega_{mn(B)}^2}\nonumber\\
& &\times N(\omega_{mn(B)})\theta(\omega_{mn(B)}).
\label{4}
\end{eqnarray}
This potential is resonant and drops as $R^{-2}$ with the distance. It means that the contribution of the resonance term due to photon absorption to the interaction potential in non-equilibrium situation could be much greater then the contribution of a non-resonance one described by the Milonni formula.
We should mention, that the approach we developed in this presentation is valid for the initial stage of interaction when the atoms and electromagnetic field are not in the equilibrium. Obviously, after the equilibration the resonance contribution to the interaction potential is zero and we obtain the Milonni formula.
 
In conclusion, we reexamined the atom-atom dispersion interaction in electromagnetic fields. We showed that the absorption and emission of virtual photons results in the resonance contributions to the non-equilibrium interaction potential.

\begin{acknowledgements}
\end{acknowledgements}

\bibliographystyle{apsrev}
\bibliography{fermometer}

\begin{thebibliography}{28}
\expandafter\ifx\csname natexlab\endcsname\relax\def\natexlab#1{#1}\fi
\expandafter\ifx\csname bibnamefont\endcsname\relax
  \def\bibnamefont#1{#1}\fi
\expandafter\ifx\csname bibfnamefont\endcsname\relax
  \def\bibfnamefont#1{#1}\fi
\expandafter\ifx\csname citenamefont\endcsname\relax
  \def\citenamefont#1{#1}\fi
\expandafter\ifx\csname url\endcsname\relax
  \def\url#1{\texttt{#1}}\fi
\expandafter\ifx\csname urlprefix\endcsname\relax\def\urlprefix{URL }\fi
\providecommand{\bibinfo}[2]{#2}
\providecommand{\eprint}[2][]{\url{#2}}

\bibitem[{\citenamefont{Mohideen and Roy}(1998)}]{Mohideen98}
\bibinfo{author}{\bibfnamefont{U.}~\bibnamefont{Mohideen}} \bibnamefont{and}
  \bibinfo{author}{\bibfnamefont{A.}~\bibnamefont{Roy}},
  \bibinfo{journal}{Phys. Rev. Lett.} \textbf{\bibinfo{volume}{81}},
  \bibinfo{pages}{4549} (\bibinfo{year}{1998}).

\bibitem[{\citenamefont{Decca et~al.}(2003)\citenamefont{Decca, L\'opez,
  Fischbach, and Krause}}]{Decca03}
\bibinfo{author}{\bibfnamefont{R.~S.} \bibnamefont{Decca}},
  \bibinfo{author}{\bibfnamefont{D.}~\bibnamefont{L\'opez}},
  \bibinfo{author}{\bibfnamefont{E.}~\bibnamefont{Fischbach}},
  \bibnamefont{and} \bibinfo{author}{\bibfnamefont{D.~E.}
  \bibnamefont{Krause}}, \bibinfo{journal}{Phys. Rev. Lett.}
  \textbf{\bibinfo{volume}{91}}, \bibinfo{pages}{050402}
  (\bibinfo{year}{2003}).

\bibitem[{\citenamefont{Bordag et~al.}(2001)\citenamefont{Bordag, Mohideen, and
  Mostepanenko}}]{Bordag01}
\bibinfo{author}{\bibfnamefont{M.}~\bibnamefont{Bordag}},
  \bibinfo{author}{\bibfnamefont{U.}~\bibnamefont{Mohideen}}, \bibnamefont{and}
  \bibinfo{author}{\bibfnamefont{V.~M.} \bibnamefont{Mostepanenko}},
  \bibinfo{journal}{Phys. Rep.} \textbf{\bibinfo{volume}{353}},
  \bibinfo{pages}{1} (\bibinfo{year}{2001}).

\bibitem[{\citenamefont{Lamoreaux}(2005)}]{Lamoreaux05}
\bibinfo{author}{\bibfnamefont{S.~K.} \bibnamefont{Lamoreaux}},
  \bibinfo{journal}{Rep. Prog. Phys.} \textbf{\bibinfo{volume}{68}},
  \bibinfo{pages}{201} (\bibinfo{year}{2005}).

\bibitem[{\citenamefont{Obrecht et~al.}(2007)\citenamefont{Obrecht, Wild,
  Antezza, Pitaevskii, Stringari, and Cornell}}]{Obrecht07}
\bibinfo{author}{\bibfnamefont{J.~M.} \bibnamefont{Obrecht}},
  \bibinfo{author}{\bibfnamefont{R.~J.} \bibnamefont{Wild}},
  \bibinfo{author}{\bibfnamefont{M.}~\bibnamefont{Antezza}},
  \bibinfo{author}{\bibfnamefont{L.~P.} \bibnamefont{Pitaevskii}},
  \bibinfo{author}{\bibfnamefont{S.}~\bibnamefont{Stringari}},
  \bibnamefont{and} \bibinfo{author}{\bibfnamefont{E.~A.}
  \bibnamefont{Cornell}}, \bibinfo{journal}{Phys. Rev. Lett.}
  \textbf{\bibinfo{volume}{98}}, \bibinfo{eid}{063201} (\bibinfo{year}{2007}).

\bibitem[{\citenamefont{Cohen and Mukamel}(2003)}]{Cohen02}
\bibinfo{author}{\bibfnamefont{A.~E.} \bibnamefont{Cohen}} \bibnamefont{and}
  \bibinfo{author}{\bibfnamefont{S.}~\bibnamefont{Mukamel}},
  \bibinfo{journal}{Phys. Rev. Lett.} \textbf{\bibinfo{volume}{91}},
  \bibinfo{pages}{233202} (\bibinfo{year}{2003}).

\bibitem[{\citenamefont{Henkel et~al.}(2002)\citenamefont{Henkel, Joulain,
  Mulet, and Greffet}}]{Henkel02}
\bibinfo{author}{\bibfnamefont{C.}~\bibnamefont{Henkel}},
  \bibinfo{author}{\bibfnamefont{K.}~\bibnamefont{Joulain}},
  \bibinfo{author}{\bibfnamefont{J.-P.} \bibnamefont{Mulet}}, \bibnamefont{and}
  \bibinfo{author}{\bibfnamefont{J.-J.} \bibnamefont{Greffet}},
  \bibinfo{journal}{J. Opt. A: Pure Appl. Opt.} \textbf{\bibinfo{volume}{4}},
  \bibinfo{pages}{S109} (\bibinfo{year}{2002}).

\bibitem[{\citenamefont{Antezza et~al.}(2005)\citenamefont{Antezza, Pitaevskii,
  and Stringari}}]{Antezza05}
\bibinfo{author}{\bibfnamefont{M.}~\bibnamefont{Antezza}},
  \bibinfo{author}{\bibfnamefont{L.~P.} \bibnamefont{Pitaevskii}},
  \bibnamefont{and}
  \bibinfo{author}{\bibfnamefont{S.}~\bibnamefont{Stringari}},
  \bibinfo{journal}{Phys. Rev. Lett.} \textbf{\bibinfo{volume}{95}},
  \bibinfo{eid}{113202} (\bibinfo{year}{2005}).

\bibitem[{\citenamefont{Antezza et~al.}(2008)\citenamefont{Antezza, Pitaevskii,
  Stringari, and Svetovoy}}]{Antezza08}
\bibinfo{author}{\bibfnamefont{M.}~\bibnamefont{Antezza}},
  \bibinfo{author}{\bibfnamefont{L.~P.} \bibnamefont{Pitaevskii}},
  \bibinfo{author}{\bibfnamefont{S.}~\bibnamefont{Stringari}},
  \bibnamefont{and} \bibinfo{author}{\bibfnamefont{V.~B.}
  \bibnamefont{Svetovoy}}, \bibinfo{journal}{Phys. Rev. A}
  \textbf{\bibinfo{volume}{77}}, \bibinfo{eid}{022901} (\bibinfo{year}{2008}).

\bibitem[{\citenamefont{Brevik et~al.}(2005)\citenamefont{Brevik, Aarseth,
  Hoye, and Milton}}]{Brevik05}
\bibinfo{author}{\bibfnamefont{I.}~\bibnamefont{Brevik}},
  \bibinfo{author}{\bibfnamefont{J.~B.} \bibnamefont{Aarseth}},
  \bibinfo{author}{\bibfnamefont{J.~S.} \bibnamefont{Hoye}}, \bibnamefont{and}
  \bibinfo{author}{\bibfnamefont{K.~A.} \bibnamefont{Milton}},
  \bibinfo{journal}{Phys. Rev. E} \textbf{\bibinfo{volume}{71}},
  \bibinfo{eid}{056101} (\bibinfo{year}{2005}).

\bibitem[{\citenamefont{Bezerra et~al.}(2006)\citenamefont{Bezerra, Decca,
  Fischbach, Geyer, Klimchitskaya, Krause, Lopez, Mostepanenko, and
  Romero}}]{Bezerra06}
\bibinfo{author}{\bibfnamefont{V.~B.} \bibnamefont{Bezerra}},
  \bibinfo{author}{\bibfnamefont{R.~S.} \bibnamefont{Decca}},
  \bibinfo{author}{\bibfnamefont{E.}~\bibnamefont{Fischbach}},
  \bibinfo{author}{\bibfnamefont{B.}~\bibnamefont{Geyer}},
  \bibinfo{author}{\bibfnamefont{G.~L.} \bibnamefont{Klimchitskaya}},
  \bibinfo{author}{\bibfnamefont{D.~E.} \bibnamefont{Krause}},
  \bibinfo{author}{\bibfnamefont{D.}~\bibnamefont{Lopez}},
  \bibinfo{author}{\bibfnamefont{V.~M.} \bibnamefont{Mostepanenko}},
  \bibnamefont{and} \bibinfo{author}{\bibfnamefont{C.}~\bibnamefont{Romero}},
  \bibinfo{journal}{Phys. Rev. E} \textbf{\bibinfo{volume}{73}},
  \bibinfo{eid}{028101} (\bibinfo{year}{2006}).

\bibitem[{\citenamefont{Mostepanenko et~al.}(2006)\citenamefont{Mostepanenko,
  Bezerra, Decca, Geyer, Fischbach, Klimchitskaya, Krause, Lopez, and
  Romero}}]{Mostepanenko06}
\bibinfo{author}{\bibfnamefont{V.~M.} \bibnamefont{Mostepanenko}},
  \bibinfo{author}{\bibfnamefont{V.~B.} \bibnamefont{Bezerra}},
  \bibinfo{author}{\bibfnamefont{R.}~\bibnamefont{Decca}},
  \bibinfo{author}{\bibfnamefont{B.}~\bibnamefont{Geyer}},
  \bibinfo{author}{\bibfnamefont{E.}~\bibnamefont{Fischbach}},
  \bibinfo{author}{\bibfnamefont{G.~L.} \bibnamefont{Klimchitskaya}},
  \bibinfo{author}{\bibfnamefont{D.~E.} \bibnamefont{Krause}},
  \bibinfo{author}{\bibfnamefont{D.}~\bibnamefont{Lopez}}, \bibnamefont{and}
  \bibinfo{author}{\bibfnamefont{C.}~\bibnamefont{Romero}},
  \bibinfo{journal}{J. Phys. A: Math. Gen.} \textbf{\bibinfo{volume}{39}},
  \bibinfo{pages}{6589} (\bibinfo{year}{2006}).

\bibitem[{\citenamefont{Chaves~de Souza~Segundo
  et~al.}(2007)\citenamefont{Chaves~de Souza~Segundo, Hamdi, Fichet, Bloch, and
  Ducloy}}]{Chaves07}
\bibinfo{author}{\bibfnamefont{P.}~\bibnamefont{Chaves~de Souza~Segundo}},
  \bibinfo{author}{\bibfnamefont{I.}~\bibnamefont{Hamdi}},
  \bibinfo{author}{\bibfnamefont{M.}~\bibnamefont{Fichet}},
  \bibinfo{author}{\bibfnamefont{D.}~\bibnamefont{Bloch}}, \bibnamefont{and}
  \bibinfo{author}{\bibfnamefont{M.}~\bibnamefont{Ducloy}},
  \bibinfo{journal}{Laser Phys.} \textbf{\bibinfo{volume}{17}},
  \bibinfo{pages}{983} (\bibinfo{year}{2007}).

\bibitem[{\citenamefont{McLachlan}(1963)}]{McLachlan63}
\bibinfo{author}{\bibfnamefont{A.~D.} \bibnamefont{McLachlan}},
  \bibinfo{journal}{Proc. R. Soc. Lond. Ser. A} \textbf{\bibinfo{volume}{274}},
  \bibinfo{pages}{80} (\bibinfo{year}{1963}).

\bibitem[{\citenamefont{Gorza and Ducloy}(2006)}]{Gorza06}
\bibinfo{author}{\bibfnamefont{M.-P.} \bibnamefont{Gorza}} \bibnamefont{and}
  \bibinfo{author}{\bibfnamefont{M.}~\bibnamefont{Ducloy}},
  \bibinfo{journal}{Eur. Phys. J. D} \textbf{\bibinfo{volume}{40}},
  \bibinfo{pages}{343} (\bibinfo{year}{2006}).

\bibitem[{\citenamefont{Buhmann and Scheel}(2008)}]{Buhmann08}
\bibinfo{author}{\bibfnamefont{S.~Y.} \bibnamefont{Buhmann}} \bibnamefont{and}
  \bibinfo{author}{\bibfnamefont{S.}~\bibnamefont{Scheel}},
  \bibinfo{journal}{arXiv:0803.0738}  (\bibinfo{year}{2008}).

\bibitem[{\citenamefont{Lifshitz}(1956)}]{Lifshitz}
\bibinfo{author}{\bibfnamefont{E.~M.} \bibnamefont{Lifshitz}},
  \bibinfo{journal}{Zh. Eksp. Teor. Fiz.} \textbf{\bibinfo{volume}{2}},
  \bibinfo{pages}{73} (\bibinfo{year}{1956}).

\bibitem[{\citenamefont{Milonni and Smith}(1996)}]{MilonniSmith96}
\bibinfo{author}{\bibfnamefont{P.~W.} \bibnamefont{Milonni}} \bibnamefont{and}
  \bibinfo{author}{\bibfnamefont{A.}~\bibnamefont{Smith}},
  \bibinfo{journal}{Phys. Rev. A} \textbf{\bibinfo{volume}{53}},
  \bibinfo{pages}{3484} (\bibinfo{year}{1996}).

\bibitem[{\citenamefont{Milonni and Shih}(1992)}]{MilonniShih92}
\bibinfo{author}{\bibfnamefont{P.~W.} \bibnamefont{Milonni}} \bibnamefont{and}
  \bibinfo{author}{\bibfnamefont{M.-L.} \bibnamefont{Shih}},
  \bibinfo{journal}{Phys. Rev. A} \textbf{\bibinfo{volume}{45}},
  \bibinfo{pages}{4241} (\bibinfo{year}{1992}).

\bibitem[{\citenamefont{Sherkunov}(2007)}]{Sherkunov07}
\bibinfo{author}{\bibfnamefont{Y.}~\bibnamefont{Sherkunov}},
  \bibinfo{journal}{Phys. Rev. A} \textbf{\bibinfo{volume}{75}},
  \bibinfo{eid}{012705} (\bibinfo{year}{2007}).

\bibitem[{\citenamefont{Power and Thirunamachandran}(1993)}]{Power93}
\bibinfo{author}{\bibfnamefont{E.~A.} \bibnamefont{Power}} \bibnamefont{and}
  \bibinfo{author}{\bibfnamefont{T.}~\bibnamefont{Thirunamachandran}},
  \bibinfo{journal}{Phys. Rev. A} \textbf{\bibinfo{volume}{47}},
  \bibinfo{pages}{2539} (\bibinfo{year}{1993}).

\bibitem[{\citenamefont{Power and Thirunamachandran}(1995)}]{Power95}
\bibinfo{author}{\bibfnamefont{E.~A.} \bibnamefont{Power}} \bibnamefont{and}
  \bibinfo{author}{\bibfnamefont{T.}~\bibnamefont{Thirunamachandran}},
  \bibinfo{journal}{Phys. Rev. A} \textbf{\bibinfo{volume}{51}},
  \bibinfo{pages}{3660} (\bibinfo{year}{1995}).

\bibitem[{\citenamefont{Casimir and Polder}(1948)}]{CasimirPolder}
\bibinfo{author}{\bibfnamefont{H.~B.~G.} \bibnamefont{Casimir}}
  \bibnamefont{and} \bibinfo{author}{\bibfnamefont{D.}~\bibnamefont{Polder}},
  \bibinfo{journal}{Phys. Rev.} \textbf{\bibinfo{volume}{73}},
  \bibinfo{pages}{360} (\bibinfo{year}{1948}).

\bibitem[{\citenamefont{Sherkunov}(2005)}]{Sherkunov05}
\bibinfo{author}{\bibfnamefont{Y.}~\bibnamefont{Sherkunov}},
  \bibinfo{journal}{Phys. Rev. A} \textbf{\bibinfo{volume}{72}},
  \bibinfo{eid}{052703} (\bibinfo{year}{2005}).

\bibitem[{\citenamefont{Keldysh}(1964)}]{Keldysh}
\bibinfo{author}{\bibfnamefont{L.}~\bibnamefont{Keldysh}},
  \bibinfo{journal}{Zh. Eksp. Teor. Fiz.} \textbf{\bibinfo{volume}{47}},
  \bibinfo{pages}{1515} (\bibinfo{year}{1964}).

\bibitem[{\citenamefont{Lifshitz and Pitaevskii}(1981)}]{LandauX}
\bibinfo{author}{\bibfnamefont{E.~M.} \bibnamefont{Lifshitz}} \bibnamefont{and}
  \bibinfo{author}{\bibfnamefont{L.~P.} \bibnamefont{Pitaevskii}},
  \emph{\bibinfo{title}{Physical kinetics. Course on theoretical physics v.10}}
  (\bibinfo{publisher}{Pergamon, Oxford}, \bibinfo{year}{1981}).

\bibitem[{\citenamefont{Wylie and Sipe}(1984)}]{WylieSipe}
\bibinfo{author}{\bibfnamefont{J.~M.} \bibnamefont{Wylie}} \bibnamefont{and}
  \bibinfo{author}{\bibfnamefont{J.~E.} \bibnamefont{Sipe}},
  \bibinfo{journal}{Phys. Rev. A} \textbf{\bibinfo{volume}{30}},
  \bibinfo{pages}{1185} (\bibinfo{year}{1984}).

\bibitem[{\citenamefont{Lifshitz and Pitaevskii}(1980)}]{LandauIX}
\bibinfo{author}{\bibfnamefont{E.~M.} \bibnamefont{Lifshitz}} \bibnamefont{and}
  \bibinfo{author}{\bibfnamefont{L.~P.} \bibnamefont{Pitaevskii}},
  \emph{\bibinfo{title}{Statistical physics, part 2, Course on theoretical
  physics v.9}} (\bibinfo{publisher}{Pergamon, Oxford}, \bibinfo{year}{1980}).

\end{thebibliography}

\end{document}